# Collaboration Tools and their Role in Agile Software Projects

Raman Mohammed Hussein[1], Bryar A. Hassan[1,2]

[1]Computer Science and Engineering Department, School of Science and Engineering, University Of Kurdistan Hawler, Erbil, Iraq.

[2]Department of Computer Science, College of Science, Charmo University, Chamchamal, Sulaimani, Iraq

Emails: raman.mohammed@ukh.edu.krd; bryar.ahmad@ukh.edu.krd

**ABSTRACT**

The purpose of this review is to understand the importance of collaboration tools which are Slack, Microsoft Teams, Confluence in Agile and software projects. Agile methodologies rely on flexibility, using cycles and integration throughout various levels of developing cycles. However, it is still a great problem for many teams to collaborate and communicate even if staff members and teams are working remotely. In terms of collaboration, the applications and technologies mean better organization of work, increased mutually understandable openness and fast and efficient inter team and interpersonal interactions to enhance results of projects into productivity. This paper examines how these tools fit the Agile principles, how they facilitate iterative development, and encouraging effective initiation and tracking of tasks in small and large projects. The insights focus on how Slack, Microsoft Teams, and Confluence are essential for gaining better task coordination, supporting knowledge sharing, and adopting agile values across cross-functional contexts.

**Introduction**

Software process improvement has always been an essential part of software projects. Current market trends and the rapid pace of changing requirements demand fast development and adaptability. Agile software development is a popular possibility to react on these trends. Implementing agile practices promises for example a shorter time-to-market, satisfied customers and increased software quality.Consequently many companies strive for an integration of agile methods or for an agile transformation.[1] In the dynamic world of software development, agile methods have become the foundation for efficiently delivering quality products and it's an increasing trend for software organizations, whether small or large, working co-located or geographically distributed [2] The foundation of success is the importance of collaboration and communication between partners. Still, there are many challenges in supporting effective collaboration and communication for agile teams [3], [4]. For example, an ever-growing plethora of different tools are needed to develop and manage software projects[5], [6], placing teams into a situation of information fragmentation and overload[6]. Iterative development, ongoing feedback, and flexible planning are key components of agile frameworks like Scrum and Kanban. Because of this, cooperation is not only advantageous but also necessary to accomplish project objectives. We'll concentrate on our well-liked products, here I'm focusing on Confluence, Microsoft Teams, and Slack. Agile teams are working differently thanks to these solutions, which provide thorough data, effective workflow management, and rapid communication. Teams may improve productivity, simplify work, and provide the finest results by utilizing these technologies. My research aims to find areas of agreement between development processes and collaboration technologies by examining the Agile work environment at three distinct sizes of companies. Therefore, I would like to respond to the following study question: How can these collaboration tools enable better communication and coordination in Agile software projects?

**Background on Agile Methodology**

Agile software development "is about feedback and change," according to Williams and Cockburn (2003) in an introduction to the IEEE Computer special issue on agile methods. They also stress that software development is an empirical or nonlinear process in which brief feedback loops are required to produce a desired, predictable result.[7] Agile Methods recognize the "need for an alternative to documentation driven, heavyweight software development processes" and are a response to conventional software development techniques.[8] Requirements "change at rates that swamp traditional methods" [9], the industry and technology are moving too quickly, and consumers are wanting more from their software while also finding it harder to clearly explain their demands up front. In order to adapt to the unavoidable shift they were going through, a number of consultants independently created procedures and methodologies.

Table 1 summarizes the key tenets of agile development, which were established by a group of software developers in 2001. Agile developers value the first item on the list more than the second in each scenario. For instance, advocates of agile approaches place a higher importance on working software than they do on documentation: We want documentation, but not hundreds of pages of infrequently used and unmaintained books.[10] [10,Beck et al., 2001].

| Table 1: Agile Principles [Beck et al., 2001] | |
|---|---|
| Agile Principle | Description |
| Customer collaboration over contract negotiation | Reduce formalities to start and finish faster, with a strong focus on the customer throughout the development process |
| Individuals and interactions over processes and tools | Enhance communication within teams and barrier removal |
| Working software over comprehensive documentation | Developers spend more time coding and testing than they do writing extensive documentation |
| Responding to change over following a plan | Give teams the freedom to make changes and adjust to project needs |

Members of cross-functional agile teams come from a variety of backgrounds, including design, testing, and development. A collaborative atmosphere where ideas and feedback are openly exchanged is fostered by seamless communication, which guarantees that everyone is in agreement. and an essential Agile value is transparency. Open sharing of updates, difficulties, and progress is facilitated by regular communication. This is made easier by tools like Microsoft Teams and Slack, which offer forums for in-the-moment communication and information exchange. Iterative cycles, or sprints, are how agile projects go forward. Every iteration grows on the one before it thanks to efficient coordination, as team members collaborate to produce small, steady improvements. Constant feedback loops are necessary for this iterative process, and effective communication techniques make this possible. Agile teams frequently have to make snap decisions in order to address problems or adjust to shifting requirements. Team members may discuss possibilities, make decisions, and execute changes immediately thanks to instant messaging and collaboration capabilities. Building trust and understanding among team members is facilitated by regular communication. A team is more cohesive and motivated when everyone feels acknowledged and appreciated, which boosts morale. Several interconnected activities are frequently included in agile projects. Dependencies are controlled and possible dangers are recognized and reduced early on thanks to effective communication. These dependencies and dangers can be tracked and documented with the use of tools such as Confluence.
Agile teams can improve their capacity to collaborate and break down communication barriers by utilizing collaboration platforms such as Slack, Microsoft Teams, and Confluence. These technologies guarantee that teams may accomplish their objectives effectively and successfully by providing the foundation required for task management, real-time discussions, and documentation.

**The Role of Collaboration Tools in Agile Projectsackground on Agile Methodology**

Team composition that promotes concentration and cooperation is beneficial for projects with high variability, such as self-organizing teams with specialized specialists. Collaboration seeks to boost output and foster innovative problem-solving. Agile teams have the option to establish collaborative teams, which can expedite the integration of various job activities, improve communication, expand knowledge sharing, and provide more assignment flexibility. Task management, real-time communication, and documentation are made possible by these technologies due to the efficiency and successfully achieved by teams through the use of these tools.

Real-time team communication is made possible by Slack, an app. It is possible to establish channels for a range of projects or topics, ensuring that conversations are easily accessible and organized. This is critical to real-time communication, which helps with agile processes such as day one stand up and decision making. "[11]

The use of Slack for coordination is facilitated by the integration of vertical, horizontal, group, and personal input channels.

The integration of Slack with other Agile project tools like Jira, Trello, and GitHub allows for the creation of channels for different projects or topics. This feature can send alerts to team members when important changes occur, such as significant deadlines.

The features of Microsoft Teams include team collaboration, video conferencing, phone calls, and various tools for managing multiple tasks. Personal interactions are made possible in remote areas through the promotion of cooperation and teamwork. Sprint reviews and retrospectives can greatly benefit from the use of calling. Teams have the option to integrate with Microsoft Planner, a visual task management tool that assists in organizing and prioritizing tasks, as well as other Microsoft Office programs like Word, Excel, and OneNote.

The use of Confluence is advantageous for coordinating, producing, and disseminating project documents. Confluence can be utilized by agile teams to store a range of information, such as meeting minutes, project plans, and technical requirements, in centralized form. It also allows multiple editors to work on the same document. Ensures that everyone has access to the latest information and can contribute to knowledge base. Facilitates the linking of documents with project tasks and issues by integrating with other collaboration platforms such as Jira and Trello.

**Detailed Overview of Each Tool**

**Slack**

A digital workspace and information management system called Slack is employed to boost team productivity. Slack transforms how companies accomplish their efficiency According to a recent survey of Slack administrators, 32.0% of participants said Slack improved their productivity by cutting meetings by 25.1% and internal emails by 48.6% [12]. Additionally. 88.6% of administrators said they felt more connected to their teams, and 79% of respondents said Slack improved the culture of their teams. When it came to information management, 80.4% of participants mentioned increased transparency in the team setting, and 62.4% said Slack made it simpler for them to find information. Slack integrates people, processes, data, agents, and artificial intelligence into a single conversational interface.

Accessibility, quicker communication, content organization, and platform connection are just a few of Slack's many helpful features. In comparison to other products, Slack was the least likely to be restricted by an institutional firewall. Members can access their workplaces through the web, a desktop plugin, or a mobile application, making it widely accessible. Additionally, Slack gives users a variety of ways to arrange content, initially by channel and then more precisely through posts and cloud-based document storage. Additionally, the technology facilitates efficient communication that goes beyond reducing email. It enables both direct, private texting and wide-ranging contact through real-time chat. All content is also archived by Slack.

Slack provides two service levels: Slack Enterprise Grid and Slack for Teams. There are three kinds of pricing for Slack for Teams: Free, Standard, and Plus. Nonprofits and educational organizations that fulfill certain requirements can receive an 85% discount on Slack's Standard and Plus services.

• Slack for Teams: Slack for Teams is designed for groups of three to one thousand individuals that share a workspace. With a basic account, Slack for Teams is free; Standard and Plus accounts cost money.

-       Free: Workspace owners can invite an infinite number of team members to join a workspace using the free version. Members of the team can share 5 GB of storage, take part in one-on-one Standard, and benefit from up to 10 third-party or custom integrations. 14 more features are available when you upgrade to a Standard account. -one video and one voice. These features include an infinite number of messages and the ability to search for individuals and channels.

-       All the capabilities of a Standard account are available to users who choose the Plus option, along with single sign-on, 99.99% informed uptime, and the ability for team owners to track and export all interactions, regardless of whether they have been deleted or are private messages.

Overall Value
Using the free version of Slack was adequate for planning a conference with a seventeen-person committee. Despite the fact that we could only connect ten Slack-approved apps, Teams can use services like Skype and Google+ Hangouts to arrange group voice and video calls, but combining too many services can be confusing and defeat the purpose of a tool designed to increase productivity, simplify communication, and centralize all data.[13]

Slack's flexibility and integration capabilities make it an excellent choice for supporting Agile methodologies. For quick updates during standups or in-depth discussions during sprint planning, Slack offers features that facilitate collaboration and communication across Agile teams.

Agile teams will appreciate this: Teams can easily navigate and communicate because to the user interface's smoothness and ease of use. Our communication processes have been considerably moved quickly by centralizing professional talks, and the availability of so many plugins, like the Giphy plugin, has given interactions a fun and innovative element.

Agile teams may lack this: It's simple to become confused and unsure of which channel is appropriate for a given topic or team when there are a lot of them. Due to the huge number of communication channels, Slack has sometimes been given less attention than other platforms, which has resulted in lower usage in some situations. The need for distinct password and email combinations for every Slack channel can seem confusing and annoying.

**Microsoft Teams**

When Microsoft announced Teams in November 2016, it entered a market firmly controlled by Slack, which had popularized business text chat as a convenient solution to email. From day one, Slack made it known it did not like Teams

Microsoft Teams is a cloud-based platform for business collaboration that provides file sharing, chat, and meetings in a shared workplace. Screen sharing, audio/video conferencing, instant messaging, and seamless integration with Microsoft 365 technologies are all supported. Typically, it is implemented throughout entire organizations rather than simply parts of it. Teams may at times use the place of email in internal business communications.

Microsoft Teams offers a range of features that enhance communication and collaboration

A whole range of capabilities are available in Microsoft Teams to improve team collaboration and communication. With continuous conversations and the option to escalate chats to audio or video calls, it facilitates both group and individual messaging. For enterprise-grade calling features, Teams integrates with the Microsoft Phone System and Calling Plan to offer cloud-based telephony as well. Advanced video conferencing tools including breakout rooms, recording, transcription, whiteboarding, and personalized backgrounds are included in virtual meetings. During calls or meetings, screen sharing enables real-time desktop sharing, and the calendar integration with Microsoft Outlook facilitates appointment and meeting management.

By March 2024, users will be able to create meeting notes and follow-up action items automatically with Microsoft Teams Premium's artificial intelligence (AI)-based features. For current Teams license holders, this add-on option provides a branded Teams Meeting feature that enables businesses to enhance invites, screens, and backdrops with colors and images to raise internal and external brand awareness.

Microsoft Teams pricing plans

Although Microsoft Teams is free to use, its subscription editions come with more capabilities and connections to other Microsoft applications. There are numerous business and enterprise plans to choose from.

Microsoft's products are aimed at a number of industry verticals. The vendor specifically targets frontline workers, healthcare, and education with Teams. There are additional one-month free trials, add-on services, nonprofit pricing, and government programs (Microsoft 365 Government G3 and G5).[14]

Teams was eliminated from Microsoft Enterprise suites worldwide on April 1, 2024, and it is no longer accessible in Office 365 E1, E3, E5, or Microsoft 365 E3 and E5. Current clients who have Office 365 or Microsoft 365 subscriptions with Teams can upgrade, renew, or change to a different licensing plan.

Overall Value

By offering a strong communication and collaboration platform, Microsoft Teams improves Agile methods. It ensures timely information and transparency by facilitating daily standups through standup bots and dedicated channels. Planning meetings, retrospectives, and continuing conversations are made easier by Teams' features including screen sharing, video conferencing, and real-time document collaboration. These features support Agile teams in staying in sync, responding quickly to problems, and continuously streamlining their operations.

Agile teams will appreciate this: They have had a really good experience with Microsoft Teams, and a few things in particular have struck out as being particularly helpful in facilitating effective teamwork and communication. It has revolutionized project collaboration by enabling smooth sharing and real-time co-editing of documents, guaranteeing that everyone is working on the most recent version.

Agile teams may lack this: Some team members can get frustrated when trying to call other Mac users, which has caused them to choose alternate means of communication. The platform has proven to have a steep learning curve for certain team members, and the organization of the interface has been viewed as unclear, which calls for attempts to make navigation easier.

**Confluence**

Confluence is a new tool that allows teams to collaborate by centralizing all of that data. Confluence arranges everything into spaces, which are groups of connected pages.
and talk about their work. Your team works in a space, which houses technical documentation, project plans and timetables, meeting notes and agendas, and more. This is your team's gateway. A small team should arrange for both team space and large-scale initiatives. We advise allocating a place for each team if you will be collaborating with many departments and teams in Confluence. The secret is to visualize a space as the container that contains everything that is significant.

The ability to create dynamic and rich Confluence pages is one of their many wonderful features. The secret ingredient that lets you get creative with your pages is macros. In Confluence, conversations take place in comments. Confluence comments appear directly next to your work, making it easier to find the criticism you require rather than having it isolated in a different system. Everyone who shares the page or finally reads it, including you as the page owner, has complete context for the conversations occurring in the comments. More than 2100 businesses worldwide currently utilize Confluence, a rapidly growing collaboration application. This technology makes cooperation easier for companies of all sizes. In addition to many SMEs, Confluence is being used by some of the biggest corporations in the world, such as Facebook, Netflix, and Stack. Because Confluence's flexibility and real-time communication tools allow development and web teams to work together agilely regardless of their geographical distance, it is quite popular among agile teams.

Agile teams will appreciate this: Many of them were positively surprised with the fact that they can use Confluence for handling documentation and problem tracking. It made it easy to use the system in handling massive docs and the overall workflow was enhanced.

Agile teams may lack this: The features that characterized Confluence that pleased the users included the documentation tool and issue tracking. The organization of its user interface was a considerable help and the program had a very good file managing factor as well.

**Survey Result**

Customers have continuously told that the benefits of Slack has given their teams and businesses since the beginning. Along with improved productivity and other advantages, they implemented things like a decrease in internal emails and meetings. they gathered thousands of survey responses from administrators and team owners of sponsored Slack teams so as to better understand the advantages. A summary of the findings is provided below.

Microsoft Teams reached 300 million users in 2023, up from 270 million it reported in 2022 Over one million organizations use Microsoft Teams as their default messaging platform, It holds 32.29% of the global video conferencing market share. 61.21% of the Microsoft Teams users are aged 35 to 54. and 47% of the medium-sized organizations use Microsoft Teams. At the same time, 32% of the large and 21% of the small organizations use Microsoft Teams. In April 2020, Microsoft Teams recorded an incredible 4.1 billion meeting minutes in just one day.[15]

| Parameters | Microsoft Teams | Slack |
|---|---|---|
| Monthly Active Users | 320 million | 54.1 million |
| Revenue | Over $8 billion | $273.4 million |

| | Number Of Organizations Using | | |
|---|---|---|---|
| Year | Slack | Microsoft teams | Confluence |
| 2020 | 750,000 | 1 million | 45842 |
| 2019 | 640,000 | 0.5 million | 38726 |
| 2018 | 450,000 | 0.2 million | 32355 |
| 2017 | 330,000 | 0.05 million | |

| Year | The Number Of Daily Active Users Of Microsoft | Number Of Daily Active Users Of Slack |
|---|---|---|
| 2023 | 280 million | 20 million |
| 2022 | 270 million | 18 million |
| 2021 | 145 million | 14 million |
| 2020 | 75 million | 12 million |
| 2019 | 20 million | 10 million |
| 2018 | 8 million | 8 million |
| 2017 | 2 million | 6 million |

**Comparison and Integration**

Slack, Microsoft Teams, and Confluence are all contouring the way businesses work with distinct advantages and disadvantages of each application. Slack is good for increasing efficiency through decreasing e-mail types and meetings, as well as improving the connections between teams and transparency. That it can be used on web, desktop and mobile is an added advantage to Agile teams especially with its integration features. Still, the problem of channel overload, the unavailability of the most options and features with the free plan, and the multi-stage sign-in are the weaknesses. Where Microsoft Teams is much more viable is that it is a full-featured chat, video call and file sharing software that is tightly integrated with Microsoft 365 and has additional features such as real-time collaboration and even an AI note-taking bot. Whereas big players can generate a great deal of value from it, adoption might entail some challenges for Teams based on the learning curve, difficulties in integrating with some Mac users, and the product being taken out of the Office 365 Enterprise packages. The fact that Confluence effectively centralizes most documentation and project management, using "spaces" and dynamic pages for problems and solutions. This need for a higher degree of documentation makes it suitable for large organizations with a lot of documentation; however, Liferay may not be good for environments that require more live communication rather than creating documentation, in addition to a has a potentially complex interface.

Since these all applications fit the particular aspect of cooperation separately, Slack, Microsoft Teams, and Confluence as one whole can theoretically build the truly synergistic collaboration. Using Slack's real-time and speedy collaboration toolbox is a great way for teams to communicate and joke around with one other. In connection with the Microsoft 365 integrated tools for file sharing and project management, Microsoft Teams can also serve as the focal point for more formal working meeting videoconferencing and document collaboration. Confluence is a wonderful fit for this course since it provides the knowledge base in the form of project timelines, documentation, and long-term storage and reference for important facts.

For instance, mentioning a Slack channel within a Confluence space will prompt notifications and vice versa. Likewise, team meetings are a great source of content that can be created immediately in Confluence and then logged there. This in turn, facilitates reconciliation of work. Additionally, syncing Teams and Slack ensures that communication keeps going smoothly without introducing new types of isolation and enables teams with individuals who prefer one medium to communicate with others who use another. Such interlinking facilitates a productive, dynamic, well-structured, and all-encompassing work environment for both current and future multi-platform integration.

**Challenges and Solutions**

People working in teams using platforms like Slack, Microsoft Teams, Confluence, etc. still face issues that affect their productivity. One of the main issues is that the number of channels, notifications, and messages makes it difficult for users to track the changes they need to make. This can be done by outlining the policies under which channels are created and encouraging the use of threads and notifications. One common problem is that different tools do not interact well with each other, resulting in silos of data that do not work well together. To combat this, these platforms need to be connected. This means that Slack can notify teams about updates in Confluence, for example, or connect to Microsoft Teams meetings and Confluence project spaces for knowledge sharing. Another downside is that if your documents are all in one place but some are spread across different platforms, you can run into sync issues and version control issues. Confluence can be used as a place to store all your documents, but Slack or Teams is your communication channel. There is also a learning curve for apps like Microsoft Teams and Confluence, which can slow adoption. To facilitate this change, it is a good idea to hold frequent sessions to provide updates and identify key players on your team to delegate platform responsibilities. Finally, working remotely is hard enough, but relationships can suffer, leading to isolation and lack of collaboration. This can be alleviated by supporting virtual team collaboration tools, supporting chat and relay, and maintaining frequent video calls to ensure team members are regularly engaged. If these issues are addressed effectively with the right approach, teams can improve communication and organizational dynamics, leading to increased productivity.

**Conclusion**

Communication tools are critical to Agile software projects because they are the primary means of interaction in an iterative setting. Slack focuses more on the communication aspect of the Agile process, while Microsoft Teams provides more information on task management and collaboration in the Agile process, while Confluence provides the documentation aspect of the Agile process. Choosing the right combination of collaboration technologies is critical to optimizing your Agile approach and keeping your team aligned, efficient, and responsive to change. By integrating these tools, organizations can make significant progress in improving their Agile projects by bridging the communication gap and sharing knowledge. Therefore, real collaboration technologies are those that enable teams to develop high-quality software that meets new and changing customer requirements.

Microsoft Teams can effortlessly organize sprints, make decisions, and stay connected across geographically dispersed locations by utilizing real-time communication solutions like Slack. Confluence, on the other hand, serves as a tool for gathering and organizing all project-related data in a way that is easy to search. This integration improves communication, leads to timely information sharing, and motivates students to keep learning and communicating.

This study demonstrates how these Agile tools can be used to improve team morale and increase productivity while reducing email and meeting communication. Additionally, teams can solve problems like notification overkill, fragmented communication, and an excessive number of tools by implementing these tools throughout Agile processes.

It is quite clear therefore that collaboration tools are not just a luxury but an essential component of supporting Agile projects, as well as helping teams self-organise, gather and respond to change and improve their processes and the quality of the software they deliver to their customers.